\documentstyle[preprint,aps]{revtex}
\begin{document}
\draft
\title{Orthorhombic distortions may reconcile\\ all the experimental
data on YBCO.}
\author{V.L.Pokrovsky}
\address{FKP-Mikrostrukturforschung,\\
ETH-Hoenggerberg, C2.2,\\
8093-Zuerich, Switzerland\cite{adad}}
\date{\today}
\maketitle
\begin{abstract}
A lot of controversy appeared recently in measurements of different
properties of High-$T_c$-superconductor YBCO. A part of data supports
d-wave hypothesis whereas other one contradicts it.
We suggest to reconcile visibly controversial experimental data
by an assumption that orthorombicity is not
small for electronic properties of YBCO, and naturally mixes s- and d-pairing.
We examine available experiments to find a proportion of such
a mixture. We find in particular that the reconcilation is plausible
if the Fermi-surface in YBCO is square-shaped.
\end{abstract}
\par
The controversy of s- and d-pairing in the High-$T_c$ superconductors
attracts much attention.
Exhaustive reviews of theoretical premises and experimental data
has been presented recently by Dynes \cite{dynes} and Schrieffer
\cite{schrieffer}.
The importance of the problem is associated with the fact that
d-pairing, if exists, would support the idea of the antiferromagnetic nature
of the Cooper interaction. In the antiferromagnetic exchange mechanism the
d-pairing occurs in a most natural way, as has been first indicated by
Scalapino {\it et al.} \cite{scalapino} and supported with detailed
calculation by Pines and coworkers \cite{pines}.
Measurements of the penetration depth \cite{hardy} and the nuclear magnetic
resonance (see discussion in \cite{schrieffer}) in YBCO have displayed
zero or very small energy gap in the excitation spectrum.
It is natural for d-pairing, and must be specially explained in the case
of s-pairing.
A mechanism of anisotropy enhancement in s-pairing has ben proposed
by Chakravarty et al.\cite{chakravarty}.
To resolve the ambiguity the experiment should answer the question,
whether the order parameter changes sign on the Fermi-surface.
Such an experiment with the Josephson tunneling in a special geometry
has been first performed by Wollman {\it et al.} \cite{wollman}.
Much improved versions of this experiment have been reported recently by
Brawner and Ott \cite{brawner} and by Mathai {\it et al.} \cite{mathai}.
The most convincing is the experiment \cite{mathai}, in which the
absence of the frozen flux and other time irreversible factors has been
reliably checked. All these experiments in our opinion prove unambigously
that the order parameter changes sign on the Fermi-surface. Together with
numerous experiments demonstrating the absence of the gap in the excitation
spectrum (\cite{schrieffer}, \cite{hardy}, \cite{valles}) they strongly
support the d-pairing idea.
\par
On the other hand, there exists a series of experiments, not less convincing
which contradicts d-pairing.
They are: i) Sun {\it et al.} \cite{sun} experiment on the Josephson tunneling
between Pb and YBCO.
If d-pairing exists, contributions to the total Josephson current from
the pairs with mutually perpendicular in-plane momenta would compensate
each other.
In the experiment \cite{sun} a non-zero Josephson current has been observed,
though rather small, 20 - 30\% of what should be expected from isotropic
superconductor; ii) Chaudhuri and Lin  \cite{lin} has applied hexagon
geometry of an YBCO sample and did not observe oscillations of current
expected at
d-pairing when switching off pairs of contact; iii) Valles {\it et al.}
\cite{Valles} and Maple {\it et al.} \cite{maple} reported the
measurements of $T_c$ vs residual resistivity $\rho$ for ion damaged
and Pr substituted YBCO respectively. Their data are reasonably close each
other and display a rather slow decrease of $T_c$ with increasing $\rho$.
The theory (\cite{radtke}, \cite{monthoux}, \cite{spokrovsky}) predicts
a strong suppression of $T_c$ by elastic impurities, which effect
d-superconductors in the same way as magnetic impurities effect
s-superconductors \cite{abrikosov}.For dirty anisotropic s-superconductors
the theory \cite{spokrovsky} predicts a power-like decrease of $T_c$ with
the residual resistivity $T_{c}(\rho)\sim\rho^{\kappa -1}$, where
$\kappa = \langle\Delta^{2}\rangle/\langle\Delta\rangle^{2} > 1$ and
$\langle...\rangle$ denotes the angular average.
It fits the experiment \cite  {sun} within the limits of experimental errors.
Kotliar and Lin \cite{kotliar} considered a mixture of s- and d-waves.
Such a mixture can appear in principle in the strong-coupling theory.
This idea has its difficulty: near the transition temperature
equations for the order parameter become
linear, if it is a second-order phase transition.
In accordance to the Landau ideology, only one order parameter, either
$d$ or $s$ can appear in the transition point. An alternative accompanying
the observation $s$ and $d$-waves together is either two subsequent
second-order phase transitions, or the first-order phase transition.
Both contradict to the experiment.
\par
The above consideration implies that the initial symmetry is tetragonal.
It is not the case for YBCO. The symmetry is almost tetragonal in CuO
planes and is obviously orthorombic in chains. In total it is orthorombic,
so that the orthorombic distortions persist in planes as well.
They do not effect much lattice constants, but can be much more substantial
for electronic properties. If they are not too large, they do not remove
the nodes, but shifted them in tetragonally asymmetric positions.
As a result $\langle\Delta\rangle\neq 0$. We show that such a picture can
reconcile  all the experimental data.
\par
Consider two simplified models.  In the first one we assume the Fermi-
surface to be a circular cylinder, and $\Delta$ to depend on the
azimuthal angle $\phi$ only:
$\Delta (\phi ) = const(\cos 2\phi + \gamma)$, where $\gamma$ is a
constant. Then $\langle\Delta\rangle = \gamma$, and
$\kappa = \langle\Delta^{2}\rangle/\langle\Delta\rangle^{2} =
 1 + 1/2\gamma^2$. In the second model the Fermi surface is a
square-shaped cylinder with the diagonals of square along $a$ and $b$-axes,
and $\Delta({\bf k}) = (\cos k_x - \cos k_y ) + \gamma$. Then again
$\langle\Delta\rangle = \gamma$ and $\langle\Delta^2\rangle = 2 +
\gamma^2$. For the value $\kappa$ in this case we get:
$\kappa = 1 + 2/\gamma^2$. From the measurements of $T_c$ vs $\rho$
 \cite{Valles}, \cite{maple} we determine $\kappa = 2.0\pm 0.3$, and
 $\gamma = 0.7\pm 0.2$ for the circular geometry, and $\gamma = 1.4 \pm
 0.2$ for the square-shaped Fermi-surface. Thus, the mixture of s-wave
 can not be small. For example, $\gamma = 0.3$ in the circular case
 gives $\kappa - 1 = 4.5$ inconsistent with the experiment. The non-zero
 average  $\langle\Delta\rangle$ explains the non-zero tunneling
 observed in the experiment \cite{sun}. It is obviously smaller than
 it could be expected for an isotropic superconductor with $\Delta^2$
 equal to $\langle\Delta^2\rangle$ in the anisotropic superconductor.
 \par
 We see that all the experiments can be reasonably described with this
 simple  sheme. To make a choice between two above described models we
 consider
 the experiment by Dolan {\it et al.} \cite{dolan}. They observed the
 anisotropy of the vortex lattice in the $a-b$-plane by the decoration
 method. The orthorhombicity is reflected in the Ginsburg-Landau tensor
 of effective masses, which is not isotropic in the orthorombic symmetry.
 We have found the ratio of two eigenvalues of effective mass tensor:
 \begin{equation}
 \frac{m_x}{m_y}\,=\,\frac{\int v_x^2\Delta^2({\bf k}) dS/v_F}
 {\int v_y^2\Delta^2({\bf k}) dS/v_F}
 \label{ratio}
 \end{equation}
 where $v_x$ and $v_y$ are components of velocity, and integration
 proceeds over the Fermi-surface. For the circular model $v_x=v_F
 \cos\phi$, $v_y=v_F \sin\phi$. The integration in (\ref{ratio})
 is straightforward with the following result:
 \begin{equation}
 \frac{m_x}{m_y}\,=\,\frac{\gamma^2 + \gamma + 1/2}
 {\gamma^2 - \gamma + 1/2}
 \label{ratiocirc}
 \end{equation}
 At $\gamma = 0.7$ this ratio takes its maximum value $\sim 5.9$.
 It means that the ratio of periods in the Abrikosov vortex lattice for the
 magnetic field along the z-axis should be $\sqrt{m_x/m_y}
 \approx 2.4$ not far from the transition temperature.
 Dolan {\it et al.} \cite{dolan}  have found for this ratio an estimate
 $\sqrt{m_x /m_y}\approx 1.15$, which is obviously incompatible with
 our estimate\cite{zhang}
For the  square-shaped Fermi-surface (\ref{ratio})
 gives the mass ratio equal to one exactly, since $v_x^2 = v_y^2$ on each side
 of the square. Therefore, experiments
 measuring $T_{c}(\rho)$ and anisotropy of effective mass together
 indicate implicitly that the shape of the Fermi-surface is close to a
 square or rectangle. We do not know a reliable experiment measuring the
Fermi-surface in YBCO. The ARPES measurements in BiSCCO \cite{dessau},
\cite{aebi}
displayed indeed the square-shaped Fermi-surface. The numerical calculations
by Massida {\it et al.} \cite{massida} also display this feature.
\par An interesting prediction which can be done from the value $\gamma=1.4$
found earlier is that the nodes of the
order parameter are located not exactly along the bisector between $a$ and
$b$-direction, but they are shifted approximately halfway to one of these axis.
 Since BiSCCO and TlBaCaCuO has tetragononal symmetry it would be very
important
 to perform tunneling experiments and measurements of $T_c$ supression
 by disorder to check whether the pure d-paring can consistently
 describe all the facts. It is interesting to note that the measurements
of the transverse magnetization in LuBaCuO \cite{buan} with the same
symmetry as YBCO displayed no nodes in the energy gap. In our language
it simply means that $\gamma > 2$.
 \par
 The experiment which still was not discussed in literature in
 connection with $s-d$ controversy is the plasma resonance. It has
 been found in YBCO by Koch  {\it et al.} \cite{koch} and in
 LaSrBaCuO by Tamasaku {\it et al.} \cite{ushida}. They observed a
 sharp decrease of reflectivity  at the plasma frequency $\Omega_{pl}$
 in superconducting state only. A theoretical analysis \cite{vpokrovsky},
 \cite{kim} shows that it possible if $\Omega_{pl}\tau < 1$.
 In both materials $\Omega_{pl}$ was slightly more than $T_{c0}$,
 the transition temperature for clean superconductors.
 It means that definitely $\tau T_c <1$.
 Such a superconductor is at least not too clean. Therefore, the energy
 gap in it  should be presumably isotropic in controversy to the
 rest of experiments. In a quite recent experiments by Harris {\it et al.}
\cite{harris} the measurement of the Hall angle have been employed
to demonstrate
that for 60K YBCO the value $\tau$ is about 2 ps, which
gives $T_c\tau\approx 16$.
 Besides of that, if there is no gap in the spectrum
 of excitations the Cooper pair  breaking is allowed kinematically
 at any frequency, and one must expect a strong plasma waves attenuation
 in the superconducting state as well. This effect can be supressed
 dynamically in clean superconductors \cite{vpokrovsky}.
\par
 Steve Kivelson kindly informed me that the idea of reconciling the
 experimental data with orthorombic symmetry of YBCO has been launched
 by V.Emery and by himself \cite{emery}. I greatly benefited discussing
this idea with V.Emery. Sharing the same general idea my work differs
substantially with the scope of details. I am indebted to H.Monien and
J.Blatter for indicating me several references and discussion.


\begin{references}
\bibitem[**]{zhang}When this work has been completed a microwave
measurement by K.Zhang {\it et al.}\cite{timusk} has been published, in which
the authors
have found for the same ratio 1.6. Though bigger than in the experiment
\cite{dolan}, this ratio is still closer to the square-shaped model value,
than to the circular one. The reason for the discrepancy in these two
experiments is not clear.
\end{references}
\end{document}